\RequirePackage{lineno}
\documentclass[preprint,tightenlines,superscriptaddress,showpacs,byrevtex]{revtex4}
\usepackage{epsfig}
\usepackage{multirow}

\usepackage{graphicx} 
\usepackage{dcolumn} 
\usepackage{rotating}
\usepackage{adjustbox}
\usepackage{threeparttable}
\usepackage[colorlinks,linkcolor=red,anchorcolor=green,citecolor=blue]{hyperref}
\usepackage{verbatim}
\usepackage{amsmath}
\usepackage{hyperref}
\usepackage[hyphenbreaks]{breakurl}

\newcommand{\BR}{{\cal B}}

\newcommand{\ks}{K_S^0}

\newcommand{\EE}{e^+e^-}

\newcommand{\beq}{\begin{equation}}
\newcommand{\eeq}{\end{equation}}
\newcommand{\bitm}{\begin{itemize}}
\newcommand{\eitm}{\end{itemize}}

\begin{document}

\title{\quad Search for $\Omega(2012)\to K\Xi(1530) \to K\pi\Xi$ at Belle}

\noaffiliation
\affiliation{University of the Basque Country UPV/EHU, 48080 Bilbao}
\affiliation{Beihang University, Beijing 100191}
\affiliation{Brookhaven National Laboratory, Upton, New York 11973}
\affiliation{Budker Institute of Nuclear Physics SB RAS, Novosibirsk 630090}
\affiliation{Faculty of Mathematics and Physics, Charles University, 121 16 Prague}
\affiliation{Chonnam National University, Kwangju 660-701}
\affiliation{University of Cincinnati, Cincinnati, Ohio 45221}
\affiliation{Deutsches Elektronen--Synchrotron, 22607 Hamburg}
\affiliation{Duke University, Durham, North Carolina 27708}
\affiliation{University of Florida, Gainesville, Florida 32611}
\affiliation{Key Laboratory of Nuclear Physics and Ion-beam Application (MOE) and Institute of Modern Physics, Fudan University, Shanghai 200443}
\affiliation{Justus-Liebig-Universit\"at Gie\ss{}en, 35392 Gie\ss{}en}
\affiliation{Gifu University, Gifu 501-1193}
\affiliation{II. Physikalisches Institut, Georg-August-Universit\"at G\"ottingen, 37073 G\"ottingen}
\affiliation{SOKENDAI (The Graduate University for Advanced Studies), Hayama 240-0193}
\affiliation{Gyeongsang National University, Chinju 660-701}
\affiliation{Hanyang University, Seoul 133-791}
\affiliation{University of Hawaii, Honolulu, Hawaii 96822}
\affiliation{High Energy Accelerator Research Organization (KEK), Tsukuba 305-0801}
\affiliation{J-PARC Branch, KEK Theory Center, High Energy Accelerator Research Organization (KEK), Tsukuba 305-0801}
\affiliation{Forschungszentrum J\"{u}lich, 52425 J\"{u}lich}
\affiliation{IKERBASQUE, Basque Foundation for Science, 48013 Bilbao}
\affiliation{Indian Institute of Science Education and Research Mohali, SAS Nagar, 140306}
\affiliation{Indian Institute of Technology Bhubaneswar, Satya Nagar 751007}
\affiliation{Indian Institute of Technology Guwahati, Assam 781039}
\affiliation{Indian Institute of Technology Hyderabad, Telangana 502285}
\affiliation{Indian Institute of Technology Madras, Chennai 600036}
\affiliation{Indiana University, Bloomington, Indiana 47408}
\affiliation{Institute of High Energy Physics, Chinese Academy of Sciences, Beijing 100049}
\affiliation{Institute of High Energy Physics, Vienna 1050}
\affiliation{Institute for High Energy Physics, Protvino 142281}
\affiliation{INFN - Sezione di Napoli, 80126 Napoli}
\affiliation{INFN - Sezione di Torino, 10125 Torino}
\affiliation{Advanced Science Research Center, Japan Atomic Energy Agency, Naka 319-1195}
\affiliation{J. Stefan Institute, 1000 Ljubljana}
\affiliation{Institut f\"ur Experimentelle Teilchenphysik, Karlsruher Institut f\"ur Technologie, 76131 Karlsruhe}
\affiliation{Kennesaw State University, Kennesaw, Georgia 30144}
\affiliation{Department of Physics, Faculty of Science, King Abdulaziz University, Jeddah 21589}
\affiliation{Kitasato University, Sagamihara 252-0373}
\affiliation{Korea Institute of Science and Technology Information, Daejeon 305-806}
\affiliation{Korea University, Seoul 136-713}
\affiliation{Kyoto University, Kyoto 606-8502}
\affiliation{Kyungpook National University, Daegu 702-701}
\affiliation{LAL, Univ. Paris-Sud, CNRS/IN2P3, Universit\'{e} Paris-Saclay, Orsay}
\affiliation{\'Ecole Polytechnique F\'ed\'erale de Lausanne (EPFL), Lausanne 1015}
\affiliation{P.N. Lebedev Physical Institute of the Russian Academy of Sciences, Moscow 119991}
\affiliation{Faculty of Mathematics and Physics, University of Ljubljana, 1000 Ljubljana}
\affiliation{Ludwig Maximilians University, 80539 Munich}
\affiliation{Luther College, Decorah, Iowa 52101}
\affiliation{University of Malaya, 50603 Kuala Lumpur}
\affiliation{University of Maribor, 2000 Maribor}
\affiliation{Max-Planck-Institut f\"ur Physik, 80805 M\"unchen}
\affiliation{School of Physics, University of Melbourne, Victoria 3010}
\affiliation{University of Mississippi, University, Mississippi 38677}
\affiliation{University of Miyazaki, Miyazaki 889-2192}
\affiliation{Moscow Physical Engineering Institute, Moscow 115409}
\affiliation{Moscow Institute of Physics and Technology, Moscow Region 141700}
\affiliation{Graduate School of Science, Nagoya University, Nagoya 464-8602}
\affiliation{Kobayashi-Maskawa Institute, Nagoya University, Nagoya 464-8602}
\affiliation{Universit\`{a} di Napoli Federico II, 80055 Napoli}
\affiliation{Nara Women's University, Nara 630-8506}
\affiliation{National Central University, Chung-li 32054}
\affiliation{National United University, Miao Li 36003}
\affiliation{Department of Physics, National Taiwan University, Taipei 10617}
\affiliation{H. Niewodniczanski Institute of Nuclear Physics, Krakow 31-342}
\affiliation{Nippon Dental University, Niigata 951-8580}
\affiliation{Niigata University, Niigata 950-2181}
\affiliation{Novosibirsk State University, Novosibirsk 630090}
\affiliation{Osaka City University, Osaka 558-8585}
\affiliation{Pacific Northwest National Laboratory, Richland, Washington 99352}
\affiliation{Peking University, Beijing 100871}
\affiliation{University of Pittsburgh, Pittsburgh, Pennsylvania 15260}
\affiliation{Punjab Agricultural University, Ludhiana 141004}
\affiliation{Research Center for Nuclear Physics, Osaka University, Osaka 567-0047}
\affiliation{Theoretical Research Division, Nishina Center, RIKEN, Saitama 351-0198}
\affiliation{University of Science and Technology of China, Hefei 230026}
\affiliation{Seoul National University, Seoul 151-742}
\affiliation{Showa Pharmaceutical University, Tokyo 194-8543}
\affiliation{Soongsil University, Seoul 156-743}
\affiliation{Stefan Meyer Institute for Subatomic Physics, Vienna 1090}
\affiliation{Sungkyunkwan University, Suwon 440-746}
\affiliation{School of Physics, University of Sydney, New South Wales 2006}
\affiliation{Department of Physics, Faculty of Science, University of Tabuk, Tabuk 71451}
\affiliation{Department of Physics, Technische Universit\"at M\"unchen, 85748 Garching}
\affiliation{Toho University, Funabashi 274-8510}
\affiliation{Department of Physics, Tohoku University, Sendai 980-8578}
\affiliation{Earthquake Research Institute, University of Tokyo, Tokyo 113-0032}
\affiliation{Department of Physics, University of Tokyo, Tokyo 113-0033}
\affiliation{Tokyo Institute of Technology, Tokyo 152-8550}
\affiliation{Tokyo Metropolitan University, Tokyo 192-0397}
\affiliation{Virginia Polytechnic Institute and State University, Blacksburg, Virginia 24061}
\affiliation{Wayne State University, Detroit, Michigan 48202}
\affiliation{Yamagata University, Yamagata 990-8560}
\affiliation{Yonsei University, Seoul 120-749}
  \author{S.~Jia}\affiliation{Beihang University, Beijing 100191} 
  \author{C.~P.~Shen}\affiliation{Key Laboratory of Nuclear Physics and Ion-beam Application (MOE) and Institute of Modern Physics, Fudan University, Shanghai 200443} 
  \author{I.~Adachi}\affiliation{High Energy Accelerator Research Organization (KEK), Tsukuba 305-0801}\affiliation{SOKENDAI (The Graduate University  for Advanced Studies), Hayama 240-0193} 
  \author{J.~K.~Ahn}\affiliation{Korea University, Seoul 136-713} 
  \author{H.~Aihara}\affiliation{Department of Physics, University of Tokyo, Tokyo 113-0033} 
  \author{S.~Al~Said}\affiliation{Department of Physics, Faculty of Science, University of Tabuk, Tabuk 71451}\affiliation{Department of Physics, Faculty of Science, King Abdulaziz University, Jeddah 21589} 
  \author{D.~M.~Asner}\affiliation{Brookhaven National Laboratory, Upton, New York 11973} 
  \author{T.~Aushev}\affiliation{Moscow Institute of Physics and Technology, Moscow Region 141700} 
  \author{R.~Ayad}\affiliation{Department of Physics, Faculty of Science, University of Tabuk, Tabuk 71451} 
  \author{V.~Babu}\affiliation{Deutsches Elektronen--Synchrotron, 22607 Hamburg} 
% ...
\collaboration{The Belle Collaboration}


\begin{abstract}

Using data samples of $e^+e^-$ collisions collected at the $\Upsilon(1S)$, $\Upsilon(2S)$, and $\Upsilon(3S)$ resonances with the Belle detector, we search for the three-body decay of the $\Omega(2012)$ baryon to $K\pi\Xi$.
This decay is predicted to dominate for models describing the $\Omega(2012)$ as a $K\Xi(1530)$ molecule. No significant $\Omega(2012)$ signals are observed in the studied channels, and 90\% credibility level upper limits on the ratios of the branching fractions relative to $K \Xi$ decay modes are obtained.

\end{abstract}


\maketitle

\section{Introduction}

Very recently a new state, the excited $\Omega(2012)$ baryon, has been observed by the Belle collaboration~\cite{121.052003} in the $\Xi K$ invariant mass spectra using data samples collected at the $\Upsilon(1S, 2S, 3S)$ energies, with measured mass
$M = [2012.4 \pm 0.7 (\rm stat.) \pm 0.6 (\rm syst.)]$~MeV/$c^{2}$ and width $\Gamma = [6.4 \pm 2.5 (\rm stat.) \pm 1.6 (\rm syst)]$~MeV.
The observed spacing in the $\Omega$ mass spectrum between the ground state and this excited state ($\sim$340 MeV/$c^2$) is smaller than that for  other $\Omega^{-}$ excited states~\cite{C38.090001}, and is more similar to the negative-parity orbital excitations of many other baryon pairs such as $\Lambda$ and $\Lambda(1405)$ or $\Lambda^{+}_{c}$ and $\Lambda^{+}_{c}(2595)$.

After the initial observation of the $\Omega(2012)$, several theoretical interpretations of that state were offered~\cite{1805.11285,1807.02145,1806.01626,1810.08318,1807.00718,1807.00997,1806.04427,1807.06485,1808.01950}. Although it is generally accepted that $\Omega(2012)$ is a $1P$ orbital excitation of the ground-state $\Omega$ baryon with quark content $sss$ and quantum numbers $J^{P}$ = $3\over 2$$^{-}$,
Refs.~\cite{1807.00997,1807.00718,1806.04427,1807.06485,1808.01950} propose an alternative interpretation as a $K\Xi(1530)$ hadronic molecule. These models predict a large decay width for $\Omega(2012) \to K\pi\Xi$. In Ref.~\cite{1807.00997}, the decay $\Omega(2012) \to K\pi\Xi$ is predicted to dominate over $\Omega(2012) \to K\Xi$, while in Refs.~\cite{1807.00718,1806.04427,1807.06485,1808.01950}, the production rates of the $\Omega(2012)$ are almost similar in $K\pi\Xi$ and $K\Xi$ decay channels.

In this paper, we report on a search for $\Omega(2012)\to K \Xi(1530) \to K\pi\Xi$ using $\Upsilon(1S,2S,3S)$
data samples collected by the Belle experiment at the KEKB asymmetric-energy $\EE$ collider~\cite{KEKB1,KEKB2}.
Note that charge-conjugate modes are implied throughout, unless explicitly stated otherwise.

\section{THE DATA SAMPLE AND BELLE DETECTOR}
The Belle data used in this analysis correspond to 5.7 fb$^{-1}$ of integrated luminosity at the $\Upsilon(1S)$ resonance, 24.9 fb$^{-1}$ at the $\Upsilon(2S)$ resonance, and 2.9 fb$^{-1}$ at the $\Upsilon(3S)$ resonance. The Belle detector~\cite{Belle1,Belle2} is a large solid-angle magnetic spectrometer consisting of a silicon vertex detector (SVD), a 50-layer central drift chamber (CDC), an array of aerogel threshold Cherenkov counters
(ACC), a barrel-like arrangement of time-of-flight scintillation counters (TOF), and an
 electromagnetic calorimeter comprised of CsI(Tl) crystals (ECL) located inside a superconducting
 solenoid coil providing a $1.5~\hbox{T}$ magnetic field. An iron flux-return yoke instrumented
 with resistive plate chambers (KLM) located outside the coil is used to detect $K^{0}_{L}$ mesons
and to identify muons. 

Large signal Monte Carlo (MC) samples (1 million events for each studied process) are generated using the {\sc Evtgen}~\cite{A462} code to simulate the expected signal event topology and estimate the signal detection efficiency. The processes $\Upsilon(1S,2S,3S)\to \Omega(2012)+anything \to K \Xi(1530) + anything \to K \pi \Xi + anything$ are simulated; the mass and width of $\Omega(2012)$ are fixed at 2.0124~GeV/$c^{2}$ and 6.4~MeV~\cite{121.052003}, respectively. To assess possible backgrounds arising from the continuum ($e^+e^- \to q \bar{q}$ with $q=u,~d,~s,~c$), we generate such events at center-of-mass energies of $\Upsilon(1S)$, $\Upsilon(2S)$, and $\Upsilon(3S)$ resonances using the Lund fragmentation model in {\sc PYTHIA}~\cite{JHEP05}. 
Inclusive $\Upsilon(1S)$ and $\Upsilon(2S)$ MC samples, corresponding to four times the luminosity of the data, are produced using {\sc PYTHIA} and are used to identify possible peaking backgrounds from $\Upsilon(1S)$ and $\Upsilon(2S)$ decays.

\section{Search for $\Omega(2012)\to K\Xi(1530) \to K\pi\Xi$}

\subsection{Event selection}

The combined information from the CDC, TOF, and ACC is used to
identify charged kaons and pions based on the kaon likelihood ratio,
$R_{K}= \mathcal{L}_{K}/(\mathcal{L}_K + \mathcal{L}_\pi)$, where $\mathcal{L}_K$
and $\mathcal{L}_{\pi}$ are the likelihood values for the kaon and pion hypotheses,
respectively.
Tracks with $R_{K}= \mathcal{L}_{K}/(\mathcal{L}_K + \mathcal{L}_\pi)<0.4$ are
identified as pions with an efficiency of 96\%, while 8\% of kaons are misidentified as pions;
tracks with $R_{K}>0.6$ are identified as kaons with an efficiency of 95\%, while 6\% of pions
are misidentified as kaons.

An ECL cluster is treated as a photon candidate if it does
not match the extrapolation of any charged track reconstructed by the tracking systems (CDC and SVD) into the calorimeter.
The $\pi^{0}$ candidates are reconstructed from two photons having energy exceeding 50~MeV in the barrel or 100~MeV in the endcaps.
To avoid contamination from neutral hadrons, we reject neutral showers if the
ratio of the energy deposited in the central array of 3$\times$3 ECL cells
to that deposited in the surrounding array of 5$\times$5 cells is less than 0.8.
The $\pi^0 \to \gamma \gamma$ candidates
are also required to have an energy balance parameter
$|E_1 - E_2|/(E_1 + E_2)$ smaller than 0.8, where $E_1$ ($E_2$) is the energy
of the first (second) photon in the laboratory frame.
To further reduce the combinatorial background, the momentum of the $\pi^{0}$ candidate is required to exceed 200 MeV/$c$. We define the $\pi^{0}$ signal region as $|M_{\gamma\gamma} - m_{\pi^{0}}|$ $<$ 12 MeV/$c^2$ ($\sim$ 2$\sigma$), where $m_{\pi^{0}}$ is the $\pi^{0}$ nominal mass~\cite{C38.090001}.
For each selected $\pi^0$ candidate, a mass-constrained fit is performed to improve its momentum resolution.

The $K_S^0$ candidates are reconstructed via the $K_S^0 \to \pi^+\pi^-$ decay, and the identification is enhanced by selecting on the outputs of a neural network~\cite{A559.190}.
The network uses the following input
variables~\cite{Thesis}: the $K_S^0$ momentum in the lab frame, the distance
along the $z$ axis between the two track helices at their
closest approach, the $K_S^0$ flight length in the $r-\phi$ plane, the
angle between the $K_S^0$ momentum and the vector joining the
interaction point (IP) to the $K_S^0$ decay vertex, the angle between the pion
momentum and the lab frame direction in the $K_S^0$ rest frame,
the distances of closest approach in the $r-\phi$ plane between the IP and the two pion helices, the number of hits in the CDC for each pion track, and the presence or absence of hits in the SVD for each pion track.

Candidate $\Lambda$ decays are reconstructed from $p\pi^-$ pairs with a production vertex significantly separated from the IP. For the $\Xi^{-}(\to\Lambda\pi^-)$ and $\Xi^{0}(\to\Lambda\pi^0)$ candidates, the vertex fits are performed and the positive $\Xi^{-}$ and $\Xi^{0}$ flight distances are required. The selected $\Xi^{-}(\to\Lambda\pi^-)$ and $\Xi^{0}(\to\Lambda\pi^0)$ candidates are the same as those in Ref.~\cite{121.052003}.
The $\Xi^{-}$ and $\Xi^{0}$ are kinematically constrained to their nominal masses~\cite{C38.090001},  and then combined with a $\pi^{\pm}$ or $\pi^{0}$ to form a $\Xi(1530)^-$ or $\Xi(1530)^0$ candidate.
Finally, the selected $\Xi(1530)$ candidate is combined with a $K^{-}$ or $K^{0}_{S}$ to form the $\Omega(2012)$ candidate.
In this last step, a vertex fit is performed for the $K\pi\Xi$ final state to improve the momentum resolutions and suppress the backgrounds, requiring $\chi^{2}_{\rm {vertex}}<20$, corresponding to an estimated selection efficiency exceeding 95\%. Reconstruction spans the $\Omega(2012)^-\to\Xi^{-}\pi^{+}K^{-}$, $\Xi^{-}\pi^{0}K^{0}_{S}$, $\Xi^{0}\pi^{-}K^{0}_{S}$, and $\Xi^{0}\pi^{0}K^{-}$ three-body decay modes of $\Omega(2012)$.

Before searching for $\Omega(2012)\to K\Xi(1530) \to K\pi\Xi$, a cross check on the previously reconstructed
$\Omega(2012)\to \Xi K$ decay mode is performed. Selection of $\Omega(2012)^-\to \Xi^{-}K^{0}_{S}/\Xi^{0}K^{-}$ candidates uses well-reconstructed  tracks, particle identifications, and vertex fitting technique in a way similar to the methods in Ref.~\cite{121.052003}.
As a result, the signal yields from the simultaneous fit of the $\Omega(2012)^-\to\Xi^{-}K^{0}_{S}$ and $\Omega(2012)^-\to\Xi^{0}K^{-}$ are $283\pm72$ and $239\pm47$, respectively.
The obtained mass and width for the $\Omega(2012)$ are $M=(2012.1\pm0.7)$ MeV/$c^{2}$ and $\Gamma=(6.9^{+2.5}_{-2.0})$ MeV, where the uncertainties are statistical only. Our results are consistent with those in Ref.~\cite{121.052003} within errors.

\subsection{The distributions from signal MC samples}

\begin{figure*}[htbp]
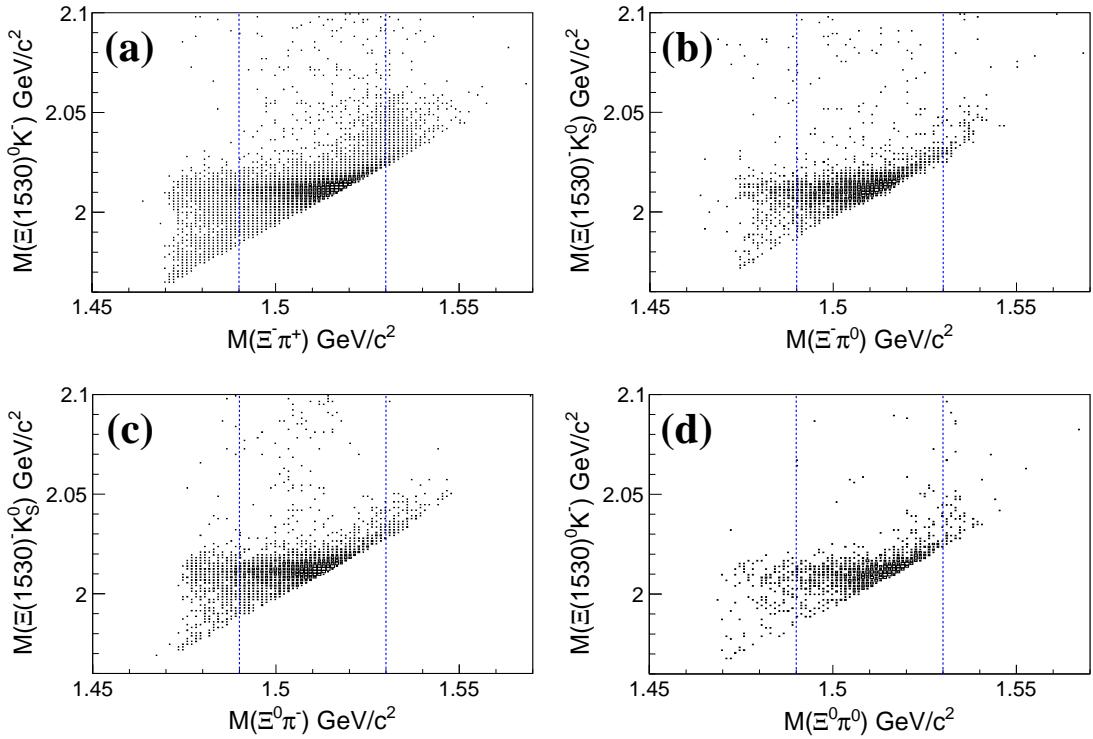

\vspace{0.15cm}
\includegraphics[height=7cm,angle=-90]{fig1a.epsi}
\hspace{0.15cm}
\vspace{0.4cm}
\includegraphics[height=7cm,angle=-90]{fig1b.epsi}

\includegraphics[height=7cm,angle=-90]{fig1c.epsi}
\hspace{0.15cm}
\includegraphics[height=7cm,angle=-90]{fig1d.epsi}
\caption{Distributions of (a) $M(\Xi(1530)^0K^-)$ versus $M(\Xi^-\pi^+)$, (a) $M(\Xi(1530)^-K^0_S)$ versus $M(\Xi^-\pi^0)$, (a) $M(\Xi(1530)^-K^0_S)$ versus $M(\Xi^0\pi^-)$, and (d) $M(\Xi(1530)^0K^-)$ versus $M(\Xi^0\pi^0)$ from signal MC samples. The dotted lines bound the $\Xi(1530)$ signal region.}\label{fig1}
\end{figure*}

After all event selection requirements, Figure~\ref{fig1} shows the distributions of the $\Xi\pi K$ invariant mass versus the $\Xi\pi$ invariant mass from signal MC samples. Due to phase space limitations, events at high $\Xi\pi$ and/or low $\Xi\pi K$ mass are kinematically forbidden.
We define the optimized $\Xi(1530)$ signal region as $1.49$ GeV/$c^{2}$ $<M(\Xi\pi)<1.53$ GeV/$c^{2}$ (discussed below), between the blue dashed lines in Fig.~\ref{fig1}.

\begin{figure*}[htbp]
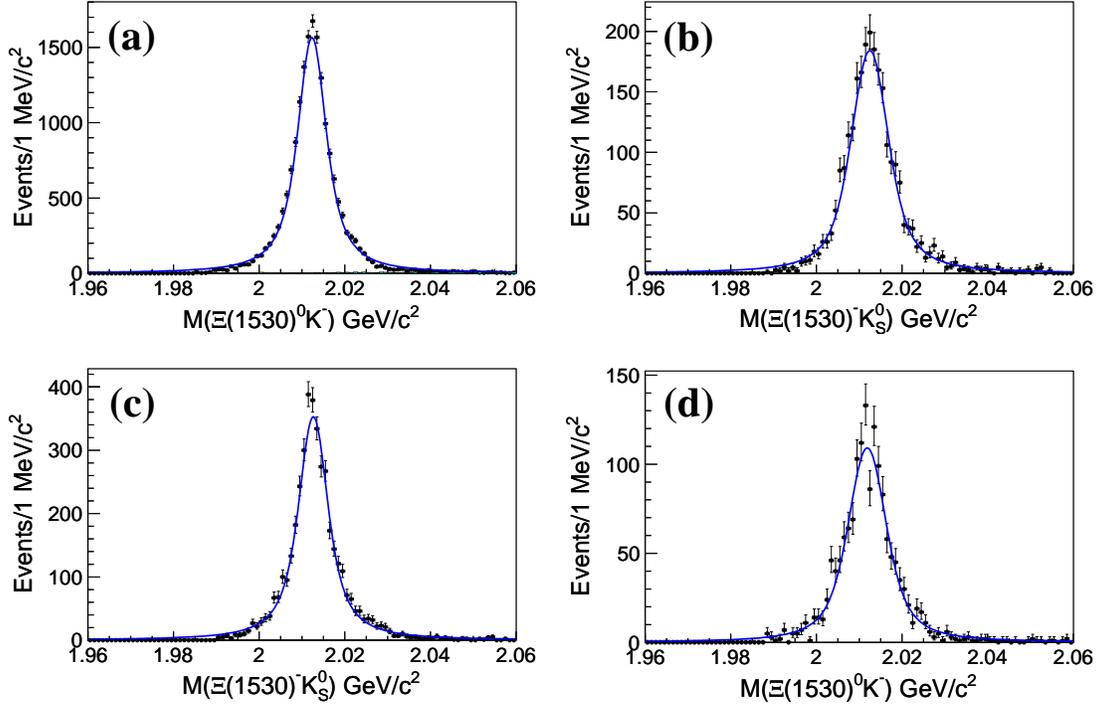

\vspace{0.15cm}
\includegraphics[height=7cm,angle=-90]{fig2a.epsi}
\hspace{0.15cm}
\vspace{0.4cm}
\includegraphics[height=7cm,angle=-90]{fig2b.epsi}

\includegraphics[height=7cm,angle=-90]{fig2c.epsi}
\hspace{0.15cm}
\includegraphics[height=7cm,angle=-90]{fig2d.epsi}
\caption{The distributions of the invariant mass of (a) $\Xi(1530)^0(\to\Xi^{-}\pi^{+})K^{-}$, (b) $\Xi(1530)^-(\to\Xi^{-}\pi^{0})K^{0}_{S}$, (c) $\Xi(1530)^-(\to\Xi^{0}\pi^{-})K^{0}_{S}$, and (d) $\Xi(1530)^0(\to\Xi^{0}\pi^{0})K^{-}$ in the signal MC samples. The solid curves show the fitted results.}\label{fig2}
\end{figure*}

The invariant mass distributions from MC signal simulations of $\Xi(1530)^0(\to\Xi^{-}\pi^{+}/\Xi^{0}\pi^{0})K^{-}$ and $\Xi(1530)^-(\to\Xi^{-}\pi^{0}/\Xi^{0}\pi^{-})K^{0}_{S}$ are shown in Fig.~\ref{fig2}.
The signal shape of the $\Omega(2012)$ is described by a Breit-Wigner (BW) function convolved with a Gaussian function, where the BW mass and width are fixed to 2.0124~GeV/$c^{2}$ and 6.4~MeV~\cite{121.052003}, respectively, and the mass-resolution Gaussian width is determined in the fit.

\subsection{$\Xi(1530)$ signals in $\Upsilon(1S,2S,3S)$ data}

\begin{figure*}[htbp]
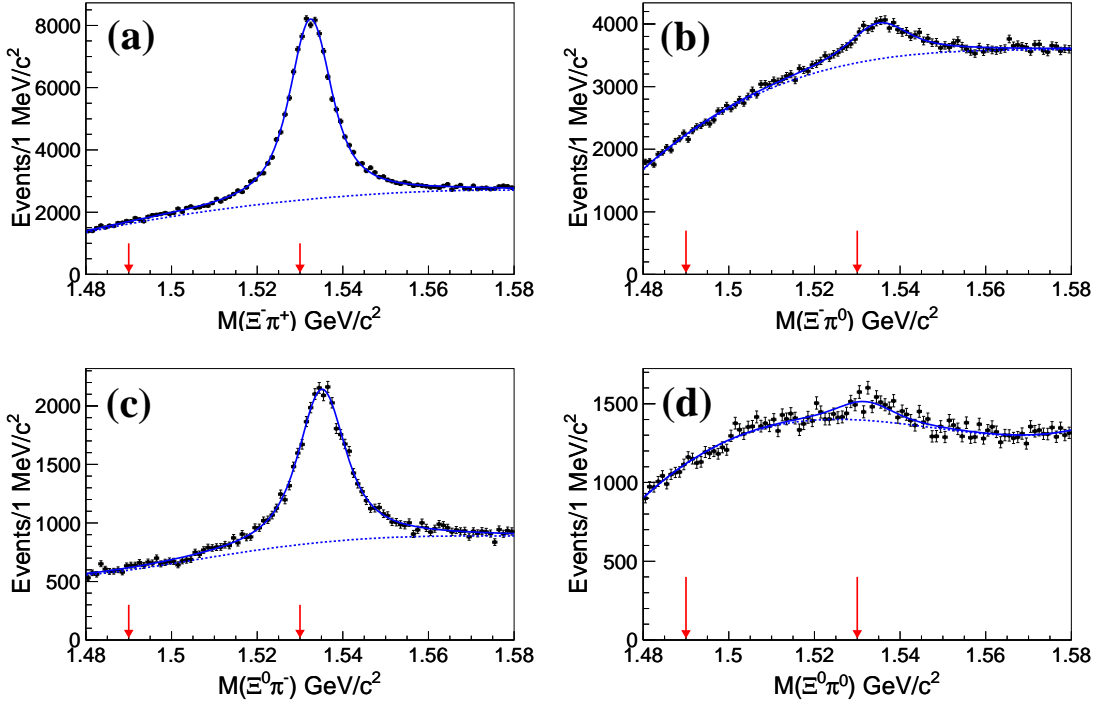

\vspace{0.15cm}
\includegraphics[height=7cm,angle=-90]{fig3a.epsi}
\hspace{0.15cm}
\vspace{0.4cm}
\includegraphics[height=7cm,angle=-90]{fig3b.epsi}

\includegraphics[height=7cm,angle=-90]{fig3c.epsi}
\hspace{0.15cm}
\includegraphics[height=7cm,angle=-90]{fig3d.epsi}
\caption{Invariant mass distributions for (a) $\Xi(1530)^0\to\Xi^{-}\pi^{+}$, (b) $\Xi(1530)^-\to\Xi^{-}\pi^{0}$, (c) $\Xi(1530)^-\to\Xi^{0}\pi^{-}$, and (d) $\Xi(1530)^0\to\Xi^{0}\pi^{0}$ candidates from the $\Upsilon(1S,2S,3S)$ data samples. Solid curves are the best fits, and dashed lines represent backgrounds. Red arrows indicate the $\Xi(1530)$ signal region for the $\Omega(2012)$ search, which is offset from the peak owing to the very limited allowed phase space.}\label{fig3}
\end{figure*}

After imposing our selection criteria, the invariant mass spectra of $\Xi(1530)^0 \to \Xi^{-}\pi^{+}, ~\Xi^{0}\pi^{0}$, and $\Xi(1530)^- \to \Xi^{-}\pi^{0}, ~\Xi^{0}\pi^{-}$ candidates are shown in Figs.~\ref{fig3}(a-d).
Clear signals of $\Xi(1530)^0$ and $\Xi(1530)^-$ are observed in the modes $\Xi(1530)^0\to\Xi^{-}\pi^{+}$ and
$\Xi(1530)^-\to\Xi^{-}\pi^{0}$, $\Xi^{0}\pi^{-}$.

We fit all the invariant mass distributions, modelling the $\Xi(1530)$ peaks with the convolution of a BW and a Gaussian function and the background as a second-order polynomial.
In the fits, the BW parameters are unconstrained, while the Gaussian widths are fixed according to MC simulations.
The fit values are consistent with the world averages within their respective errors~\cite{C38.090001}.
For $\Xi(1530)^0\to\Xi^{0}\pi^{0}$,  the mass and width of $\Xi(1530)^0$ are fixed
to the Particle Data Group (PDG) values~\cite{C38.090001} since the signal is not clear due to large combinatorial backgrounds.
The results of the fits are listed in Table~\ref{1530}.

\begin{table*}[htbp]
\vspace{0.01cm}
\caption{Mass resolution from MC simulations, and the mass and width for inclusive $\Xi(1530)$ signals from fits to the Belle data.}\label{1530}
\begin{tabular}{c | c | c | c}
\hline\hline
~Mode~ & ~~Resolution~(MeV/$c^{2}$)~~ & ~~Mass~(MeV/$c^{2}$)~~ & ~~Width~(MeV)~~ \\\hline
$\Xi(1530)^0\to\Xi^{-}\pi^{+}$ &$2.34\pm0.14$ &$1532.47\pm0.03$ &$9.0\pm0.3$ \\
$\Xi(1530)^-\to\Xi^{-}\pi^{0}$ &$2.96\pm0.17$ &$1535.07\pm0.37$ &$12.9\pm1.8$ \\
$\Xi(1530)^-\to\Xi^{0}\pi^{-}$ &$2.44\pm0.15$ &$1535.11\pm0.09$ &$10.6\pm0.2$ \\
$\Xi(1530)^0\to\Xi^{0}\pi^{0}$ &$4.14\pm0.26$ & 1531.80 (PDG value)&9.1 (PDG value) \\
\hline\hline
\end{tabular}
\end{table*}

\subsection{$\Omega(2012)\to\Xi \pi K$ mass distributions in $\Upsilon(1S,2S,3S)$ data}

Considering phase space limitations and our finite mass resolution, we require $1.49$ GeV/$c^{2}$ $<M(\Xi\pi)<1.53$ GeV/$c^{2}$ to select $\Xi(1530)$ signals as efficiently as possible, as indicated by the red arrows in Fig.~\ref{fig3}.
We optimize this requirement by maximizing the figure of merit $N_{sig}/\sqrt{N_{sig}+N_{bkg}}$ value with the mode $\Omega(2012)^-\to\Xi(1530)^0(\to\Xi^-\pi^+)K^-$, where $N_{sig}$ is number of fitted signal events in the signal MC sample assuming $\BR(\Upsilon(1S,2S,3S)\to\Omega(2012)^- +
{\rm anything})\times\BR(\Omega(2012)^-\to\Xi(1530)^0K^{-})$ = 10$^{-6}$ and $N_{bkg}$ is the number of estimated background events in the $\Omega(2012)^-$ signal region using inclusive MC samples.
The candidate signal region for the $\Xi(1530)$ coincides with the predicted mass interval from Ref.~\cite{1807.06485}.

\begin{figure*}[htbp]
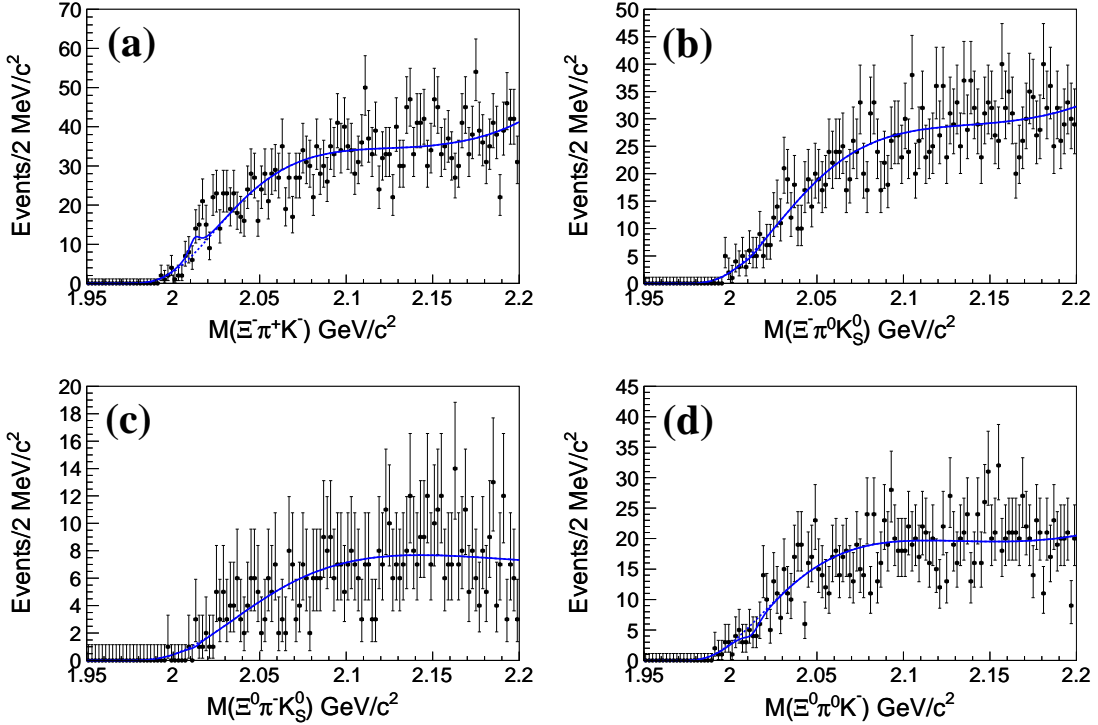

\vspace{0.15cm}
\includegraphics[height=7cm,angle=-90]{fig4a.epsi}
\hspace{0.15cm}
\vspace{0.4cm}
\includegraphics[height=7cm,angle=-90]{fig4b.epsi}

\includegraphics[height=7cm,angle=-90]{fig4c.epsi}
\hspace{0.15cm}
\includegraphics[height=7cm,angle=-90]{fig4d.epsi}
\caption{The distributions of the invariant mass for (a) $\Xi(1530)^0(\to\Xi^{-}\pi^{+})K^{-}$, (b) $\Xi(1530)^-(\to\Xi^{-}\pi^{0})K^{0}_{S}$, (c) $\Xi(1530)^-(\to\Xi^{0}\pi^{-})K^{0}_{S}$, and (d) $\Xi(1530)^0(\to\Xi^{0}\pi^{0})K^{-}$ from the $\Upsilon(1S,2S,3S)$ data samples. The solid curves are the best fits, and the dashed lines represent the backgrounds.}\label{fig4}
\end{figure*}

After application of the above selection criteria, Fig.~\ref{fig4} shows the invariant mass distributions of $\Xi(1530)^0(\to\Xi^{-}\pi^{+}/\Xi^{0}\pi^{0})K^{-}$ and $\Xi(1530)^-(\to\Xi^{-}\pi^{0}/\Xi^{0}\pi^{-})K^{0}_{S}$. From these distributions, no obvious $\Omega(2012)^-$ signal is observed. The shapes of the $\Omega(2012)$ signals in the fits are described by BW functions convolved with Gaussian resolution functions; the background shapes are described by a threshold function.
The parameters of the BW functions are fixed to the mass and width of the $\Omega(2012)$~\cite{121.052003}, and the mass resolutions are fixed
to those from fits to signal MC samples (1.5, 2.6, 1.7, and 2.8~MeV
for the $\Omega(2012)\to \Xi(1530)^0(\to\Xi^{-}\pi^{+})K^{-}$,
$\Xi(1530)^-(\to\Xi^{-}\pi^{0})K^{0}_{S}$, $\Xi(1530)^-(\to\Xi^{0}\pi^{-})K^{0}_{S}$, and $\Xi(1530)^0(\to\Xi^{0}\pi^{0})K^{-}$
decay modes, respectively). The threshold function has the form $(M(\Xi K)-x)^{\alpha}$exp$[c_{1}(M(\Xi K)-x)+c_{2}(M(\Xi K)-x)^{2}]$, where the parameters $\alpha$, $c_{1}$, and $c_{2}$  are free; the threshold parameter $x$ is fixed at 1.97~GeV/$c^{2}$ from the MC simulations.
The yields of $\Omega(2012)$ signal events from the unbinned extended maximum-likelihood fits are obtained; they are listed in Table~\ref{results}, together with the reconstruction efficiency, signal significance, and the upper limit at 90\% credibility level~\cite{CL} (C.L.) on the signal yield for each $\Omega(2012)$ decay mode.
In addition, no peaking backgrounds are found from the inclusive MC samples.

\begin{table*}[htbp]
\vspace{0.01cm}
\caption{The reconstruction efficiency ($\varepsilon$), signal significance ($\sigma$), signal yield ($N^{\rm fit}$), and the upper limit at 90\% C.L. ($N^{\rm UL}$) on the signal yield for each $\Omega(2012)$ decay mode.}\label{results}
\begin{tabular}{c | c | c | c | c}
\hline\hline
~~~Mode~~~ & ~~~~~$\varepsilon$~(\%)~~~~~ & ~~~$\sigma$~~~ & ~~~~~$N^{\rm fit}$~~~~~ & ~~~~$N^{\rm UL}$~~~~ \\\hline
$\Omega(2012)^-\to \Xi(1530)^0(\to\Xi^{-}\pi^{+})K^{-}$ &$8.71\pm0.06$ &1.8&$22.5\pm12.9$ &41.0 \\
$\Omega(2012)^-\to \Xi(1530)^-(\to\Xi^{-}\pi^{0})K^{0}_{S}$ &$1.26\pm0.01$ &-&$-3.5\pm11.6$ &16.6 \\
$\Omega(2012)^-\to \Xi(1530)^-(\to\Xi^{0}\pi^{-})K^{0}_{S}$ &$2.06\pm0.02$ &-&$-1.0\pm3.6$ &7.2 \\
$\Omega(2012)^-\to \Xi(1530)^0(\to\Xi^{0}\pi^{0})K^{-}$ &$0.75\pm0.01$ &-&$-12.0\pm9.8$ &13.2 \\
\hline\hline
\end{tabular}
\end{table*}

\subsection{The ratios of the branching fractions for $\Omega(2012)\to K\pi\Xi$ relative to $K\Xi$}

We define the ratios $\mathcal{R}^{\Xi^{-}\pi^{+}K^{-}}_{\Xi^{-}\bar K^{0}}$, $\mathcal{R}^{\Xi^{-}\pi^{0}\bar K^{0}}_{\Xi^{-}\bar K^{0}}$, $\mathcal{R}^{\Xi^{0}\pi^{-}\bar K^{0}}_{\Xi^{0}K^{-}}$, $\mathcal{R}^{\Xi^{-}\pi^{+}K^{-}}_{\Xi^{0}K^{-}}$, $\mathcal{R}^{\Xi^{0}\pi^{-}\bar K^{0}}_{\Xi^{-}\bar K^{0}}$, and $\mathcal{R}^{\Xi^{0}\pi^{0}K^{-}}_{\Xi^{0}K^{-}}$ and determine their values as follows:
\begin{equation} \label{eq:1}
\scriptsize
\mathcal{R}^{\Xi^{-}\pi^{+}K^{-}}_{\Xi^{-}\bar K^{0}}=\frac{ \BR(\Omega(2012)\to \Xi(1530)^0(\to \Xi^{-}\pi^{+})K^{-})} {\BR(\Omega(2012)\to \Xi^{-}\bar K^{0})} = \frac{N^{\rm fit}_{1}\times \varepsilon_{5} \times \BR(K^{0}_{S}\to \pi^{+}\pi^{-}) \times\BR(\bar K^{0} \to K^{0}_{S})}{N^{\rm fit}_{5}\times \varepsilon_{1}},
\end{equation}
\begin{equation} \label{eq:2}
\scriptsize
\mathcal{R}^{\Xi^{-}\pi^{0}\bar K^{0}}_{\Xi^{-}\bar K^{0}}=\frac{ \BR(\Omega(2012)\to \Xi(1530)^-(\to\Xi^{-}\pi^{0})\bar K^{0})} {\BR(\Omega(2012)\to \Xi^{-}\bar K^{0})} = \frac{N^{\rm fit}_{2}\times \varepsilon_{5}}{N^{\rm fit}_{5}\times \varepsilon_{2} \times \BR(\pi^{0}\to \gamma\gamma)},
\end{equation}
\begin{equation} \label{eq:3}
\scriptsize
\mathcal{R}^{\Xi^{0}\pi^{-}\bar K^{0}}_{\Xi^{0}K^{-}}=\frac{ \BR(\Omega(2012)\to \Xi(1530)^-(\to \Xi^{0}\pi^{-})\bar K^{0})} {\BR(\Omega(2012)\to \Xi^{0}K^{-})} = \frac{N^{\rm fit}_{3}\times \varepsilon_{6} }{N^{\rm fit}_{6}\times \varepsilon_{3} \times \BR(K^{0}_{S}\to \pi^{+}\pi^{-}) \times\BR(\bar K^{0} \to K^{0}_{S})},
\end{equation}
\begin{equation} \label{eq:4}
\scriptsize
\mathcal{R}^{\Xi^{0}\pi^{0}K^{-}}_{\Xi^{0}K^{-}}=\frac{ \BR(\Omega(2012)\to \Xi(1530)^0(\to\Xi^{0}\pi^{0})K^{-})} {\BR(\Omega(2012)\to \Xi^{0}K^{-})} = \frac{N^{\rm fit}_{4}\times \varepsilon_{6} }{N^{\rm fit}_{6}\times \varepsilon_{4} \times \BR(\pi^{0}\to \gamma\gamma)},
\end{equation}
\begin{equation} \label{eq:10}
\scriptsize
\mathcal{R}^{\Xi^{-}\pi^{+}K^{-}}_{\Xi^{0}K^{-}}=\frac{ \BR(\Omega(2012)\to \Xi(1530)^0(\to \Xi^{-}\pi^{+})K^{-})} {\BR(\Omega(2012)\to \Xi^{0}K^{-})} = \frac{N^{\rm fit}_{1}\times \varepsilon_{6} \times \BR(\Xi^{0}\to \Lambda\pi^{0})\times\BR(\pi^0\to\gamma\gamma)}{N^{\rm fit}_{6}\times \varepsilon_{1}\times \BR(\Xi^{-}\to \Lambda\pi^{-})},
\end{equation}
\begin{equation} \label{eq:11}
\scriptsize
\mathcal{R}^{\Xi^{0}\pi^{-}\bar K^{0}}_{\Xi^{-}\bar K^{0}}=\frac{ \BR(\Omega(2012)\to \Xi(1530)^-(\to \Xi^{0}\pi^{-})\bar K^{0})} {\BR(\Omega(2012)\to \Xi^{-}\bar K^{0})} = \frac{N^{\rm fit}_{3}\times \varepsilon_{5} \times \BR(\Xi^{-}\to \Lambda\pi^{-})}{N^{\rm fit}_{5}\times \varepsilon_{3}\times \BR(\Xi^{0}\to \Lambda\pi^{0})\times \BR(\pi^0\to\gamma\gamma)},
\end{equation}
where the errors are statistical only; $N^{\rm fit}_{1}$,
$N^{\rm fit}_{2}$,
$N^{\rm fit}_{3}$,
$N^{\rm fit}_{4}$,
$N^{\rm fit}_{5}$, and
$N^{\rm fit}_{6}$ are the fitted signal yields in the modes
$\Omega(2012)^-\to \Xi(1530)^0(\to \Xi^{-}\pi^{+})K^{-}$,
$\Xi(1530)^-(\to\Xi^{-}\pi^{0})\bar K^{0}$,
$\Xi(1530)^-(\to \Xi^{0}\pi^{-})\bar K^{0}$,
$\Xi(1530)^0(\to\Xi^{0}\pi^{0})K^{-}$,
$\Xi^{-}K^{0}_{S}$, and $\Xi^{0} K^-$,
respectively; $\varepsilon_{1}$,
$\varepsilon_{2}$,
$\varepsilon_{3}$,
$\varepsilon_{4}$,
$\varepsilon_{5}$, and
$\varepsilon_{6}$ are the corresponding
efficiencies for each mode. The values of $N^{\rm fit}_{1}$, $N^{\rm fit}_{2}$, $N^{\rm fit}_{3}$, $N^{\rm fit}_{4}$, $\varepsilon_{1}$, $\varepsilon_{2}$, $\varepsilon_{3}$, and $\varepsilon_{4}$ are listed in Table~\ref{results}. The values of $N^{\rm fit}_{5}$, $N^{\rm fit}_{6}$, $\varepsilon_{5}$, and $\varepsilon_{6}$ are $279\pm71$, $242\pm48$, $(15.7\pm0.2)$\%, and $(4.0\pm0.1)$\%.
In our calculations, we use the standard value of $\BR(\bar K^{0} \to K^{0}_{S})$ = 0.5.
Finally, the values of the $\mathcal{R}^{\Xi^{-}\pi^{+}K^{-}}_{\Xi^{-}\bar K^{0}}$, $\mathcal{R}^{\Xi^{-}\pi^{0}\bar K^{0}}_{\Xi^{-}\bar K^{0}}$, $\mathcal{R}^{\Xi^{0}\pi^{-}\bar K^{0}}_{\Xi^{0}K^{-}}$, $\mathcal{R}^{\Xi^{0}\pi^{0}K^{-}}_{\Xi^{0}K^{-}}$, $\mathcal{R}^{\Xi^{-}\pi^{+}K^{-}}_{\Xi^{0}K^{-}}$, and $\mathcal{R}^{\Xi^{0}\pi^{-}\bar K^{0}}_{\Xi^{-}\bar K^{0}}$ are obtained; they are listed in Table~\ref{yields}.

\begin{table*}[htbp]
\vspace{0.01cm}
\caption{The values of the $\mathcal{R}^{\Xi^{-}\pi^{+}K^{-}}_{\Xi^{-}\bar K^{0}}$, $\mathcal{R}^{\Xi^{-}\pi^{0}\bar K^{0}}_{\Xi^{-}\bar K^{0}}$, $\mathcal{R}^{\Xi^{0}\pi^{-}\bar K^{0}}_{\Xi^{0}K^{-}}$, $\mathcal{R}^{\Xi^{0}\pi^{0}K^{-}}_{\Xi^{0}K^{-}}$, $\mathcal{R}^{\Xi^{-}\pi^{+}K^{-}}_{\Xi^{0}K^{-}}$, and $\mathcal{R}^{\Xi^{0}\pi^{-}\bar K^{0}}_{\Xi^{-}\bar K^{0}}$.}\label{yields}
\begin{tabular}{c | c }
\hline\hline
~~~The ratio~~~ & ~~~~~The value~~~~~ \\\hline
$\mathcal{R}^{\Xi^{-}\pi^{+}K^{-}}_{\Xi^{-}\bar K^{0}}$&$(5.0\pm2.9)$\% \\
$\mathcal{R}^{\Xi^{-}\pi^{0}\bar K^{0}}_{\Xi^{-}\bar K^{0}}$&$(-15.8\pm52.3)$\% \\
$\mathcal{R}^{\Xi^{0}\pi^{-}\bar K^{0}}_{\Xi^{0}K^{-}}$&$(-2.3\pm8.4)$\% \\
$\mathcal{R}^{\Xi^{0}\pi^{0}K^{-}}_{\Xi^{0}K^{-}}$&$(-26.8\pm21.9)$\% \\
$\mathcal{R}^{\Xi^{-}\pi^{+}K^{-}}_{\Xi^{0}K^{-}}$&$(4.2\pm2.5)$\% \\
$\mathcal{R}^{\Xi^{0}\pi^{-}\bar K^{0}}_{\Xi^{-}\bar K^{0}}$&$(-2.8\pm10.0)$\% \\
\hline\hline
\end{tabular}
\end{table*}

\subsection{Simultaneous fit results}

Considering that the branching fractions of $\Omega(2012)^-\to \Xi^{-}\bar K^{0}$ and $\Omega(2012)^-\to \Xi^{0}K^{-}$ and
the ratios of branching fractions of the three-body decay modes of $\Omega(2012)$ are known, the ratio of expected signal yields between each $\Omega(2012)$ three-body decay mode can be calculated. With such constraints, we perform a simultaneous fit to obtain the upper limit on $\mathcal{R}^{\Xi\pi K}_{\Xi K}$ = $\BR(\Omega(2012)\to \Xi(1530)(\to \Xi\pi )K)/\BR(\Omega(2012)\to \Xi K)$.

Taking $\BR(\Omega(2012)^-\to \Xi^{-}\bar K^{0}):\BR(\Omega(2012)^-\to \Xi^{0}K^{-})$ = $1.0:1.2$~\cite{121.052003} and $\BR(\Omega(2012)^-\to \Xi(1530)^0(\to \Xi^{-}\pi^{+})K^{-}) :\BR(\Omega(2012)^-\to \Xi(1530)^-(\to\Xi^{-}\pi^{0})\bar K^{0}) : \BR(\Omega(2012)^-\to \Xi(1530)^-(\to \Xi^{0}\pi^{-})\bar K^{0}): \BR(\Omega(2012)^-\to \Xi(1530)^0(\to\Xi^{0}\pi^{0})K^{-})$ = $2:1:2:1$ according to isospin symmetry, we derive that $\mathcal{R}^{\Xi^{-}\pi^{+}K^{-}}_{\Xi^{-}\bar K^{0}}$: $\mathcal{R}^{\Xi^{-}\pi^{0}\bar K^{0}}_{\Xi^{-}\bar K^{0}}$: $\mathcal{R}^{\Xi^{0}\pi^{-}\bar K^{0}}_{\Xi^{0}K^{-}}$: $\mathcal{R}^{\Xi^{0}\pi^{0}K^{-}}_{\Xi^{0}K^{-}}$=$1:\frac{1}{2}:\frac{1}{1.2}:\frac{1}{2.4}$. Thus, according to Eqs.~(\ref{eq:1}--\ref{eq:4}), we have:
\begin{equation} \label{eq:5}
N^{\rm fit}_{1} : N^{\rm fit}_{2} : N^{\rm fit}_{3} : N^{\rm fit}_{4} = 87.2\% : 2.2\% : 7.0\% : 3.6\%.
\end{equation}

\begin{figure*}[htbp]
\vspace{0.3cm}
\includegraphics[height=9cm,angle=-90]{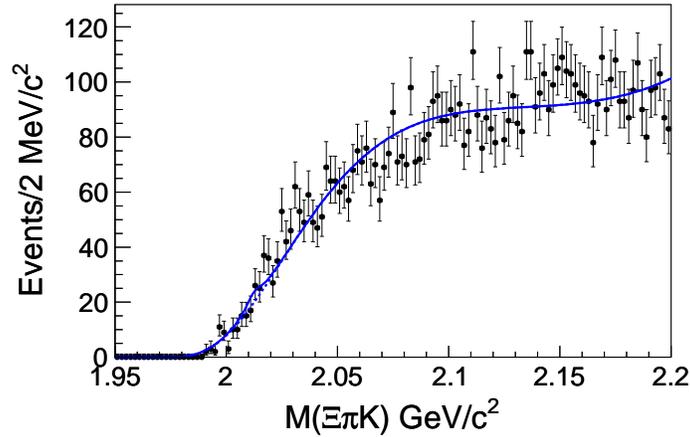}
\caption{The final simultaneous fit result to all three-body $\Omega(2012)$ decay modes from the combined $\Upsilon(1S,2S,3S)$ data samples. The solid curve is the best fit, and the dashed line represents the backgrounds.}\label{fig5}
\end{figure*}

An unbinned extended maximum-likelihood simultaneous fit to all three-body decay modes is now performed. In the simultaneous fit, the ratios of the expected observed $\Omega(2012)$ signals between each decay channel are fixed according to Eq.~(\ref{eq:5}).
The functions used to describe the signal and background shapes are parameterized as before.
The fit result is shown in Fig.~\ref{fig5} from the combined $\Upsilon(1S,2S,3S)$ data samples, corresponding to a total fit yield of $22.4\pm14.0$.  The statistical significance of the $\Omega(2012)$ signal is $1.6\sigma$. Finally, we determine
\begin{equation} \label{eq:14}
\mathcal{R}^{\Xi\pi K}_{\Xi K}=\frac{\BR(\Omega(2012)\to \Xi(1530)(\to \Xi\pi )K)} {\BR(\Omega(2012)\to \Xi K)} = (6.0\pm3.7({\rm stat.})\pm1.3({\rm syst.}))\%,
\end{equation}
where $\BR(\Omega(2012)\to \Xi(1530)(\to \Xi\pi)K$) = $\BR(\Omega(2012)^-\to \Xi(1530)^0(\to \Xi^{-}\pi^{+})K^{-})$ + $\BR(\Omega(2012)^-\to \Xi(1530)^-(\to\Xi^{-}\pi^{0})\bar K^{0})$ + $\BR(\Omega(2012)^-\to \Xi(1530)^-(\to \Xi^{0}\pi^{-})\bar K^{0})$ + $\BR(\Omega(2012)^-\to \Xi(1530)^0(\to\Xi^{0}\pi^{0})K^{-})$ and $\BR(\Omega(2012)\to \Xi K)$ = $\BR(\Omega(2012)^-\to \Xi^{-}\bar K^{0})$ + $\BR(\Omega(2012)^-\to \Xi^{0}K^{-})$. In the calculations, each branching fraction is determined individually.
Systematic uncertainties are detailed below.

\subsection{Systematic uncertainties}

We now discuss the systematic uncertainties inherent in our measurements of the ratios $\mathcal{R}^{\Xi^{-}\pi^{+}K^{-}}_{\Xi^{-}\bar K^{0}}$, $\mathcal{R}^{\Xi^{-}\pi^{0}\bar K^{0}}_{\Xi^{-}\bar K^{0}}$, $\mathcal{R}^{\Xi^{0}\pi^{-}\bar K^{0}}_{\Xi^{0}K^{-}}$, $\mathcal{R}^{\Xi^{0}\pi^{0}K^{-}}_{\Xi^{0}K^{-}}$, $\mathcal{R}^{\Xi^{-}\pi^{+}K^{-}}_{\Xi^{0}K^{-}}$, $\mathcal{R}^{\Xi^{0}\pi^{-}\bar K^{0}}_{\Xi^{-}\bar K^{0}}$, and $\mathcal{R}^{\Xi\pi K}_{\Xi K}$. These include detection efficiency (tracking efficiency, kaon and pion particle ID, $\Lambda$, $\ks$, and $\pi^{0}$ reconstruction), the statistical error in the MC efficiency, the branching fractions of possible intermediate states, the $\Omega(2012)$ resonance parameters, any possible bias in reconstructed mass (as evaluated from the difference between the reconstructed $\Xi^{0}$ mass and the world average value), as well as the overall fit uncertainty.

Based on a study of $D^{*+}\to D^{0}(\to K^0_S\pi^+\pi^-)\pi^+$, the uncertainty in tracking efficiency is taken to be 0.35\% per track.
The uncertainties in particle identification are studied via a low-background sample of $D^{*}$ decay for charged kaons and pions. The studies show uncertainties of 1.3\% for each charged kaon and 1.1\% for each charged pion. The uncertainty in $\Lambda$ selection is 3\%~\cite{D94.032002}. Differences in $K_S^0$ selection efficiency determined from data and MC simulation give a relation of $1 - \varepsilon_{\rm data}/\varepsilon_{\rm MC} = (1.4 \pm 0.3)\%$~\cite{119.171801}; 1.7\% is taken as a conservative systematic uncertainty. For $\pi^{0}$ reconstruction, the efficiency correction and systematic uncertainty are estimated from a sample of $\tau^{-}\to\pi^-\pi^0\nu$.
We find a 2.25\% systematic uncertainty on $\pi^0$ reconstruction efficiency.
In the measurements of $\mathcal{R}^{\Xi^{-}\pi^{+}K^{-}}_{\Xi^{-}\bar K^{0}}$, $\mathcal{R}^{\Xi^{-}\pi^{0}\bar K^{0}}_{\Xi^{-}\bar K^{0}}$, $\mathcal{R}^{\Xi^{0}\pi^{-}\bar K^{0}}_{\Xi^{0}K^{-}}$, $\mathcal{R}^{\Xi^{-}\pi^{+}K^{-}}_{\Xi^{0}K^{-}}$, $\mathcal{R}^{\Xi^{0}\pi^{-}\bar K^{0}}_{\Xi^{-}\bar K^{0}}$, and $\mathcal{R}^{\Xi^{0}\pi^{0}K^{-}}_{\Xi^{0}K^{-}}$, the common sources of systematic uncertainties such as $\Xi$ selection cancel; the individual errors are summed in quadrature to obtain the total detection efficiency uncertainty.
For the measurement of $\mathcal{R}^{\Xi\pi K}_{\Xi K}$, to determine the total detection efficiency, the systematic errors for each final state and the errors from tracking, particle identification, $\Lambda$, $\ks$, and $\pi^{0}$ reconstruction are first summed in quadrature to obtain $\sigma_{i}$. Then, the total systematic uncertainty for detection efficiency ($\sigma_{\rm DE}$) is determined using standard error propagation as follows:
\begin{equation}
\sigma_{\rm DE} = \sqrt{\frac{\Sigma_{i}({\cal{W}}_{i}\times \sigma_{i})^{2}}{(\Sigma_{i}{\cal{W}}_{i})^{2}}+\frac{\Sigma_{j}({\cal{W}}_{j}\times \sigma_{j})^{2}}{\Sigma_{j}({\cal{W}}_{j})^{2}}} = 7.3\%
\end{equation}
Here, ${\cal{W}}_{i}$ is the weight factor of the branching fraction in the $i$-th ($i$ = 0, 1, 2, 3) mode of the $\Omega(2012)\to \Xi\pi K$ decay; ${\cal{W}}_{j}$ ($j$ = 0, 1) is the relative weight for the $j$-th mode of $\Omega(2012)\to \Xi K $ decay.

The statistical uncertainty in the determination of the efficiency from MC simulations is less than 1.0\%.
In the calculation of $\mathcal{R}^{\Xi\pi K}_{\Xi K}$, only the branching fractions of intermediate states $\BR(K^{0}_{S}\to \pi^{+}\pi^{-})$ and $\BR(\pi^{0}\to \gamma\gamma)$ are included; the corresponding uncertainties are 0.072\% and 0.035\%~\cite{C38.090001}, respectively, which are sufficiently small to be neglected.
The uncertainty in the $\Omega(2012)$ resonance parameters is estimated by toggling the values of
resonance mass and width by $\pm$1$\sigma$ and refitting.
The largest differences compared to the nominal fit results are taken as the systematic uncertainties associated with the $\Omega(2012)$ resonance parameters. The uncertainty in the $\Xi^{0}$ mass is estimated by comparing the numbers of the signal yields of the $\Omega(2012)$ for the case where the mass of the reconstructed $\Xi^{0}$ is fixed at the found peak value versus the case where the mass is fixed to the nominal mass~\cite{C38.090001}.
According to the $\Xi(1530)K$ invariant mass distributions in inclusive MC samples, we find that the threshold mass value falls within the [1.96, 1.98]~GeV/$c^{2}$ interval. The systematic error in the background parameterization is estimated by comparing the yields when the threshold mass is changed by $\pm$10 MeV/$c^{2}$ relative to the nominal fit (for which the threshold is fixed at 1.97 GeV/$c^{2}$).

All the uncertainties are summarized in Table~\ref{systematic}, and, assuming all errors are independent, summed in quadrature to give the total systematic uncertainty.

\begin{table}[htbp]
\caption{Relative systematic errors (\%) on the measurements of $\mathcal{R}^{\Xi^{-}\pi^{+}K^{-}}_{\Xi^{-}\bar K^{0}}$, $\mathcal{R}^{\Xi^{-}\pi^{0}\bar K^{0}}_{\Xi^{-}\bar K^{0}}$, $\mathcal{R}^{\Xi^{0}\pi^{-}\bar K^{0}}_{\Xi^{0}K^{-}}$, $\mathcal{R}^{\Xi^{0}\pi^{0}K^{-}}_{\Xi^{0}K^{-}}$, $\mathcal{R}^{\Xi^{-}\pi^{+}K^{-}}_{\Xi^{0}K^{-}}$, $\mathcal{R}^{\Xi^{0}\pi^{-}\bar K^{0}}_{\Xi^{-}\bar K^{0}}$, and $\mathcal{R}^{\Xi\pi K}_{\Xi K}$.}\label{systematic}
\begin{tabular}{c | c  c  c  c  c c c } \hline
Source & $\mathcal{R}^{\Xi^{-}\pi^{+}K^{-}}_{\Xi^{-}\bar K^{0}}$ & $\mathcal{R}^{\Xi^{-}\pi^{0}\bar K^{0}}_{\Xi^{-}\bar K^{0}}$ & $\mathcal{R}^{\Xi^{0}\pi^{-}\bar K^{0}}_{\Xi^{0}K^{-}}$ & $\mathcal{R}^{\Xi^{0}\pi^{0}K^{-}}_{\Xi^{0}K^{-}}$ &  $\mathcal{R}^{\Xi^{-}\pi^{+}K^{-}}_{\Xi^{0}K^{-}}$&  $\mathcal{R}^{\Xi^{0}\pi^{-}\bar K^{0}}_{\Xi^{-}\bar K^{0}}$& $\mathcal{R}^{\Xi\pi K}_{\Xi K}$ \\\hline
Detection efficiency & 2.5& 3.4& 2.6& 3.0&3.3&3.3& 7.3\\
MC statistics & 1.0& 1.0& 1.0& 1.0&1.0&1.0& 1.0\\
$\Omega(2012)$ resonance parameters & 10.7 & 33.5& 41.3& 27.8&10.7&41.3& 6.1  \\
$\Xi^{0}$ mass & -& -& 17.4& 3.3&-&17.4& 4.5\\
Background parameter & 7.9& 23.4& 30.0& 17.2&7.9&30.0& 18.1\\\hline
Sum in quadrature &13.6 & 41.0& 54.0& 33.0&13.7&54.1& 21.0\\\hline
\end{tabular}
\end{table}

\subsection{90\% C.L. upper limits}

In the absence of any significant observed signals, upper limits at 90\% C.L. on the $\mathcal{R}^{\Xi^{-}\pi^{+}K^{-}}_{\Xi^{-}\bar K^{0}}$, $\mathcal{R}^{\Xi^{-}\pi^{0}\bar K^{0}}_{\Xi^{-}\bar K^{0}}$, $\mathcal{R}^{\Xi^{0}\pi^{-}\bar K^{0}}_{\Xi^{0}K^{-}}$, $\mathcal{R}^{\Xi^{0}\pi^{0}K^{-}}_{\Xi^{0}K^{-}}$, $\mathcal{R}^{\Xi^{-}\pi^{+}K^{-}}_{\Xi^{0}K^{-}}$, $\mathcal{R}^{\Xi^{0}\pi^{-}\bar K^{0}}_{\Xi^{-}\bar K^{0}}$, and $\mathcal{R}^{\Xi\pi K}_{\Xi K}$ modes are determined by solving the equation
\begin{equation} \label{eq:UL}
\int^{t^{\rm UL}}_0 \mathcal{F}_{\rm likelihood}(t)dt / \int^{+\infty}_0\mathcal{F}_{\rm likelihood}(t)dt = 90\%,
\end{equation}
where $t$ is the
assumed ratio of branching fractions, and $\mathcal{F}_{\rm likelihood}(t)$ is the corresponding maximized likelihood of the data. To take into account systematic uncertainties, the likelihood is convolved with a Gaussian function whose width equals the corresponding total systematic uncertainty. Finally, we obtain
\begin{equation}  \label{eq:final1}
\mathcal{R}^{\Xi^{-}\pi^{+}K^{-}}_{\Xi^{-}\bar K^{0}}=\frac{ \BR(\Omega(2012)\to \Xi(1530)^0(\to \Xi^{-}\pi^{+})K^{-})} {\BR(\Omega(2012)\to \Xi^{-}\bar K^{0})} < 9.3\%,
\end{equation}
\begin{equation} \label{eq:final2}
\mathcal{R}^{\Xi^{-}\pi^{0}\bar K^{0}}_{\Xi^{-}\bar K^{0}}=\frac{ \BR(\Omega(2012)\to \Xi(1530)^-(\to\Xi^{-}\pi^{0})\bar K^{0})} {\BR(\Omega(2012)\to \Xi^{-}\bar K^{0})} < 81.1\%,
\end{equation}
\begin{equation} \label{eq:final3}
\mathcal{R}^{\Xi^{0}\pi^{-}\bar K^{0}}_{\Xi^{0}K^{-}}=\frac{ \BR(\Omega(2012)\to \Xi(1530)^-(\to \Xi^{0}\pi^{-})\bar K^{0})} {\BR(\Omega(2012)\to \Xi^{0}K^{-})} < 21.3\%,
\end{equation}
\begin{equation} \label{eq:final4}
\mathcal{R}^{\Xi^{0}\pi^{0}K^{-}}_{\Xi^{0}K^{-}}=\frac{ \BR(\Omega(2012)\to \Xi(1530)^0(\to\Xi^{0}\pi^{0})K^{-})} {\BR(\Omega(2012)\to \Xi^{0}K^{-})} < 30.4\%,
\end{equation}
\begin{equation} \label{eq:12}
\mathcal{R}^{\Xi^{-}\pi^{+}K^{-}}_{\Xi^{0}K^{-}}=\frac{ \BR(\Omega(2012)\to \Xi(1530)^0(\to \Xi^{-}\pi^{+})K^{-})} {\BR(\Omega(2012)\to \Xi^{0}K^{-})} < 7.8\%,
\end{equation}
\begin{equation} \label{eq:13}
\mathcal{R}^{\Xi^{0}\pi^{-}\bar K^{0}}_{\Xi^{-}\bar K^{0}}=\frac{ \BR(\Omega(2012)\to \Xi(1530)^-(\to \Xi^{0}\pi^{-})\bar K^{0})} {\BR(\Omega(2012)\to \Xi^{-}\bar K^{0})} < 25.6\%,
\end{equation}
and
\begin{equation}  \label{eq:final5}
\mathcal{R}^{\Xi\pi K}_{\Xi K}=\frac{ \BR(\Omega(2012)\to \Xi(1530)(\to\Xi\pi) K)} {\BR(\Omega(2012)\to \Xi K)} < 11.9\%
\end{equation}
at 90\% C.L.

\section{RESULTS AND DISCUSSION}

In summary, using the data samples of 5.7 fb$^{-1}$ $\Upsilon(1S)$, 24.9 fb$^{-1}$ $\Upsilon(2S)$, and 2.9 fb$^{-1}$ $\Upsilon(3S)$ collected by the Belle detector, we have searched for the three-body~$K\pi\Xi$ decay of $\Omega(2012)$ for the first time.
No significant signals are observed, and we determine upper limits at 90\% C.L. on the ratios of $\mathcal{R}^{\Xi^{-}\pi^{+}K^{-}}_{\Xi^{-}\bar K^{0}}$, $\mathcal{R}^{\Xi^{-}\pi^{0}\bar K^{0}}_{\Xi^{-}\bar K^{0}}$, $\mathcal{R}^{\Xi^{0}\pi^{-}\bar K^{0}}_{\Xi^{0}K^{-}}$, $\mathcal{R}^{\Xi^{0}\pi^{0}K^{-}}_{\Xi^{0}K^{-}}$, $\mathcal{R}^{\Xi^{-}\pi^{+}K^{-}}_{\Xi^{0}K^{-}}$, $\mathcal{R}^{\Xi^{0}\pi^{-}\bar K^{0}}_{\Xi^{-}\bar K^{0}}$, and $\mathcal{R}^{\Xi\pi K}_{\Xi K}$ to be 9.3\%, 81.1\%, 21.3\%, 30.4\%, 7.8\%, 25.6\%, and 11.9\%, respectively.
Our result strongly disfavors the
molecular interpretation proposed by Ref.~\cite{1807.00997}, and is in tension with the predictions of Refs.~\cite{1807.00718,1806.04427,1807.06485,1808.01950}, also based on molecular interpretations.

\section{ACKNOWLEDGMENTS}

S. Jia acknowledges the support from the China Scholarship Council under No. 201706020133 and the Academic Excellence Foundation of BUAA for PhD students.
We thank Professor Li-sheng Geng for useful discussions and comments.
We thank the KEKB group for the excellent operation of the
accelerator; the KEK cryogenics group for the efficient
operation of the solenoid; and the KEK computer group, and the Pacific Northwest National
Laboratory (PNNL) Environmental Molecular Sciences Laboratory (EMSL)
computing group for strong computing support; and the National
Institute of Informatics, and Science Information NETwork 5 (SINET5) for
valuable network support.  We acknowledge support from
the Ministry of Education, Culture, Sports, Science, and
Technology (MEXT) of Japan, the Japan Society for the
Promotion of Science (JSPS), and the Tau-Lepton Physics
Research Center of Nagoya University;
the Australian Research Council including grants
DP180102629, 
DP170102389, 
DP170102204, 
DP150103061, 
FT130100303; 
Austrian Science Fund (FWF);
the National Natural Science Foundation of China under Contracts
No.~11435013,  
No.~11475187,  
No.~11521505,  
No.~11575017,  
No.~11675166,  
No.~11705209;  
Key Research Program of Frontier Sciences, Chinese Academy of Sciences (CAS), Grant No.~QYZDJ-SSW-SLH011; 
the  CAS Center for Excellence in Particle Physics (CCEPP); 
the Shanghai Pujiang Program under Grant No.~18PJ1401000;  
the Ministry of Education, Youth and Sports of the Czech
Republic under Contract No.~LTT17020;
the Carl Zeiss Foundation, the Deutsche Forschungsgemeinschaft, the
Excellence Cluster Universe, and the VolkswagenStiftung;
the Department of Science and Technology of India;
the Istituto Nazionale di Fisica Nucleare of Italy;
National Research Foundation (NRF) of Korea Grants
No.~2015H1A2A1033649, No.~2016R1D1A1B01010135, No.~2016K1A3A7A09005
603, No.~2016R1D1A1B02012900, No.~2018R1A2B3003 643,
No.~2018R1A6A1A06024970, No.~2018R1D1 A1B07047294; Radiation Science Research Institute, Foreign Large-size Research Facility Application Supporting project, the Global Science Experimental Data Hub Center of the Korea Institute of Science and Technology Information and KREONET/GLORIAD;
the Polish Ministry of Science and Higher Education and
the National Science Center;
the Grant of the Russian Federation Government, Agreement No.~14.W03.31.0026; 
the Slovenian Research Agency;
Ikerbasque, Basque Foundation for Science, Spain;
the Swiss National Science Foundation;
the Ministry of Education and the Ministry of Science and Technology of Taiwan;
and the United States Department of Energy and the National Science Foundation.


\begin{thebibliography}{**}
\bibitem{CL} In common high energy physics usage, this Bayesian interval has been reported as ``confidence interval,'' which is a frequentist-statistics term.
\end{thebibliography}
\end{document}